\documentclass[aps,preprint,superscriptaddress,nofootinbib]{revtex4}%
\usepackage{amsfonts}
\usepackage{amsmath}
\usepackage{amssymb}
\usepackage{graphicx}%

\newcommand{\be}{\begin{equation}}
\newcommand{\ee}{\end{equation}}
\newcommand{\bea}{\begin{eqnarray}}
\newcommand{\eea}{\end{eqnarray}}

\begin{document}

\title[Short title for running header]
    {\textbf{CANONICAL ANALYSIS OF THE BCEA TOPOLOGICAL MATTER MODEL COUPLED TO
             GRAVITATION IN (2+1) DIMENSIONS}}

\author{\textbf{Laurent Freidel}}

    \email{lfreidel@perimeterinstitute.ca}

    \affiliation{Laboratoire de Physique, {\'E}cole Normale Sup{\'e}rieure de Lyon,
                 46 All{\'e}e d'Italie, 69364 Lyon Cedex 07, France\\}

    \affiliation{Perimeter Institute for Theoretical Physics, Waterloo,
                 Ontario, Canada, N2J 2G9}

\author{\textbf{R.\ B.\ Mann}}

    \email{mann@avatar.uwaterloo.ca}

    \affiliation{Perimeter Institute for Theoretical Physics, Waterloo,
                 Ontario, Canada, N2J 2G9}

    \affiliation{Department of Physics, University of Waterloo, Waterloo,
                 Ontario, Canada N2L 3G1}

\author{\textbf{Eugeniu M.\ Popescu}}

    \email{empopesc@uwaterloo.ca}

    \affiliation{Department of Physics, University of Waterloo, Waterloo,
                 Ontario, Canada N2L 3G1}

\begin{abstract}

    \noindent
        We consider a topological field theory derived from the Chern -
        Simons action in (2+1) dimensions with the $I(ISO(2,1))$ group, and
        we investigate in detail the canonical structure of this theory.
        Originally developed as a  topological theory of Einstein gravity
        minimally coupled to topological matter fields in (2+1) dimensions,
        it admits a BTZ black-hole solutions, and can be generalized to
        arbitrary dimensions. In this paper, we further study the canonical structure
        of the theory in (2+1) dimensions, by identifying all the distinct
        gauge equivalence classes of solutions as they result from
        holonomy considerations. The equivalence classes are discussed in
        detail, and examples of solutions representative of each class are
        constructed or identified.
\end{abstract}

\date[Date: ]{\today}

\maketitle

\section{Introduction}

\indent
    Because of the difficulties of quantizing gravity in (3+1)
    dimensions, general relativity in (2+1)-dimensional spacetimes
    emerged as a lower-dimensional alternative, whose purpose
    was to help in understanding at least part of the issues
    involved in the development of a quantum theory of gravitation.\\
\indent
    Much simpler than its (3+1)-dimensional counterpart since it
    has no propagating modes, general relativity in (2+1) dimensions
    in the absence of matter was shown by Witten \cite{Witten1} to
    be exactly solvable. If matter is added via the coupling of gravity
    to pointlike particles solvability is preserved (see \cite{FL}
    for a recent review on the quantization of this system). Unfortunately,
    when matter is added in the traditional way, by coupling gravity to a
    field theory, solvability is, even for this simple theory, generally
    destroyed.\\
\indent
    It would appear from the above considerations that the
    addition of matter automatically destroys the solvability of pure
    general relativity, and while this statement is indeed true in
    many instances and for various dimensions of spacetimes, in (2+1)
    dimensions there is a notable exception. In order to understand
    this aspect, which is characteristic to (2+1)-dimensional general
    relativity, it is necessary to make a short digression into how
    matter is coupled to pure general relativity.\\
\indent
    In the Lagrangian formalism, the Hilbert-Einstein action for pure
    gravity is a functional of a single variable \cite{Wald}, the
    spacetime metric, and it is through the spacetime metric that
    matter is usually coupled to the theory. In doing so, the space of
    states of the theory becomes infinite-dimensional, and the
    solvability of the theory is destroyed. In the Palatini formalism,
    however, the action functional for pure gravity is considered to
    be a functional not of a single variable, as in the
    Hilbert-Einstein case, but of two variables, the spacetime metric
    and the components of the connection (alternatively, the spacetime
    metric and the spacetime derivatives of the metric), which are now
    considered to be independent variables. Correspondingly, in the
    Palatini formalism there are two possibilities for coupling
    matter to pure gravity: through the metric, and through the
    connection. Of course, coupling matter through the metric is no
    different from coupling matter in the Hilbert-Einstein formalism,
    leading to the same solvability issues, just in a different
    framework.\\
\indent
    However, coupling of matter through the connection yields
    significantly different results. In this latter context, there is
    a distinct class of topological field theories where the matter
    fields introduce only a finite number of degrees of freedom, such
    that the total space of states remains finite-dimensional,
    rendering these theories solvable both classically and quantum
    mechanically. It is this particular class of topological field
    theories that constitutes the notable exception to the
    (in)solvability vs. matter issue mentioned earlier.\\
\indent
    In the present paper, we consider one particular such topological
    matter model which was originally developed by Carlip and
    Gegenberg \cite{bcea}. The model (subsequently referred to as the
    BCEA model) consists of a pair of 1-form matter fields $B$, $C$
    that are minimally coupled to the first order action of pure
    gravity through the connection 1-form fields via the covariant
    derivative (see next section), and as mentioned earlier, it is
    exactly solvable both classically and quantum mechanically. What
    makes this model interesting is the fact that it is non-trivial
    for non-trivial topologies of the spacetime foliation. In
    particular, if the spacelike leaves of the foliation have the
    topology of a plane with one puncture , it was shown
    \cite{bceabtz} that the model admits a solution that is analogous
    to the BTZ black hole, and this fact suggests the possibility that
    at least for this particular case, the model could have a much
    richer structure. Based on this observation, we have decided, as
    part of a larger ongoing project, to investigate in detail the
    general classical structure of the BCEA model in the case where
    the topology of the spacelike foliation is that of a plane with
    one puncture, in order to determine all the distinct gauge
    equivalence classes of solution.\\
\indent
    The paper is organized as follows. In Section II we present
    a short summary of the important features of the model relevant
    for our discussion. In Section III, based on holonomy
    considerations, we perform the canonical analysis of the
    structure of the model for the case where the leaves of the
    foliation have the topology of the punctured plane, and we identify
    all the the distinct sectors of the theory. In section IV,
    we illustrate the different sectors of the theory by
    constructing or identifying solutions that are
    of physical relevance, and in Section V we conclude with some
    remarks and considerations regarding future work.\\

\section{The BCEA theory}

\indent
    The action of the BCEA model in the first order formalism
    has the expression:
        \be
            S[B,C,E,A]=\int_{M}(E_{i}\wedge R^{i}[A]+B_{i}\wedge DC^{i})\label{sbcea}
        \ee
    where $M$ is a 3-dimensional non-compact spacetime with the the topology $M$=%
    $R\times \Sigma $, and $\Sigma $ is a 2-dimensional spacelike surface with
    the topology of a plane with one puncture. The fields $E_{i}$ in (\ref{sbcea})
    are $SO(2,1)$ 1-forms which, if invertible, correspond to the triads of
    the spacetime metric, and $R^{i}[A]$ are the curvature 2-forms associated with
    the $SO(2,1)$ connection 1-forms $A^{i}$, with the expression:
        \be
            R^{i}[A]=dA^{i}+\frac{1}{2}\,\epsilon ^{ijk}A_{j}\wedge A^{k}\label{RA}
        \ee
    The $SO(2,1)$ 1-forms $B^{i}$, $C^{i}$, are the topological matter fields
    that are coupled to the fields $E^{i}$, $A^{i}$ of pure gravity,
    and $DC^{i}$ is the covariant derivative of the field $C^{i}$,
    having the expression:
        \be
            DC^{i}=dC^{i}+\,\epsilon ^{ijk}A_{j}\wedge C_{k}\label{covder}
        \ee
    Throughout the entire paper we adopt the following index convention. Greek
    indices, taking the values $0$, $1$, $2$, designate the spacetime components
    of tensors, and are raised and lowered by the spacetime metric $g_{\alpha
    \beta }$. Latin lower case indices,also taking the values $0$, $1$, $2$,
    are $SO(2,1)$ indices,and are raised and lowered by the $SO(2,1)$ metric $
    \eta _{ij}$=$diag(-1,1,1)$, and $\epsilon ^{ijk}$ is the totally
    antisymmetric $SO(2,1)$ symbol with $\epsilon ^{012}$=$1$.\\
\indent
    The action (\ref{sbcea}) yields, upon first order variation
    (and up to surface terms), the equations of motion:
        \bea
            R^{i}[A]&=&0  \nonumber \\
            DE^{i}+\,\epsilon ^{ijk}B_{j}\wedge C_{k}&=&0 \\
            DB^{i}=DC^{i}&=&0  \nonumber  \label{eqnmotn}
        \eea
    and is invariant under the following 12-parameter infinitesimal gauge
    transformations:\newline
        \bea
            \delta A^{i}&=&D\tau ^{i}\nonumber\\
            \delta B^{i}&=&D\rho ^{i}+\,\epsilon ^{ijk}B_{j}\tau _{k}\nonumber\\
            \delta C^{i}&=&D\lambda ^{i}+\,\epsilon ^{ijk}C_{j}\tau _{k}\\
            \delta E^{i}&=&D\beta ^{i}+\,\epsilon ^{ijk}(E_{j}\tau _{k}+
                            B_{j}\lambda_{k}+C_{j}\rho _{k})\nonumber\label{eqngauge}
        \eea
    where $\beta ^{i}$, $\lambda ^{i}$, $\rho ^{i}$, $\tau ^{i}$ are 0-form
    gauge parameters.\\
\indent
    The $(2+1)$ canonical splitting induced by the topology of
    the manifold $M$ yields four sets of constraints ${J^{i}}$,
    ${P^{i}}$, ${Q^{i}}$, ${R^{i}}$, which are enforced by the
    zeroth spacetime components of the form fields $A^{i}$, $E^{i}$,
    $B^{i}$, and $C^{i}$ respectively, acting as Lagrange multipliers.
    The Lie algebra generated by these constraints is:
        \bea
            \{J^{i},J^{j}\} &=&\epsilon ^{ijk}J_{k}  \nonumber \\
            \{J^{i},P^{j}\} &=&\epsilon ^{ijk}P_{k}  \nonumber \\
            \{J^{i},Q^{j}\} &=&\epsilon ^{ijk}Q_{k}  \label{alg} \\
            \{J^{i},R^{j}\} &=&\epsilon ^{ijk}R_{k}  \nonumber \\
            \{Q^{i},R^{j}\} &=&\epsilon ^{ijk}P_{k}  \nonumber
        \eea
    with the rest of the Poisson brackets being zero. The algebra
    (\ref{alg}) is the inhomogeneization of the Poincar\'{e} algebra
    determined by $\{J^{i},Q^{i}\}$ with the abelian generators
    $\{P^{i},R^{i}\}$, and as such it can be recognized as the Lie
    algebra of the inhomogeneized Poincar\'{e} group $I(ISO(2,1))$
    \cite{GKL}. The Hamiltonian of the system is zero on shell, since
    it depends only on the constraints, and consequently the constraints are
    preserved in time.\\
\indent
    On the Lie algebra of $I(ISO(2,1))$ we can introduce an
    invariant scalar product\footnote[1]{The most general scalar product that
    one can introduce can also contain terms of the form
    $\tilde{tr}(J_{i}Q_{j})=\alpha \eta _{ij}$, $\tilde{tr}(J_{i}R_{j})=\gamma \eta _{ij}$.
    One can however redefine the algebra generators such that the algebra is
    left invariant by this redefinition, and recover the scalar product
    (\ref{scalprod})} for the generators defined as:
        \bea
            tr(J_{i}P_{j})=tr(Q_{i}R_{j})=\eta _{ij}  \label{scalprod}
        \eea
    with all the other pairings being zero, and a generalized connection
    ${\cal{A}}$ having the expression:
        \be
            {\cal {A}}=A_{i}J^{i}+E_{i}P^{i}+B_{i}Q^{i}+C_{i}R^{i}\label{gencon}
        \ee
    With these definitions, it is straightforward to show that the action (\ref%
    {sbcea}) of the BCEA model can be written, up to surface terms, as a
    Chern-Simons theory with the connection ${\cal {A}}$ and the invariant
    scalar product (\ref{scalprod}):
        \be
            S[B,C,E,A]=\frac{1}{2}\,\int_{M}tr({\cal {A}}\wedge d{\cal {A}}+
            \frac{2}{3}\,{\cal {A}}\wedge {\cal {A}}\wedge {\cal {A}})
        \ee
    Moreover, and in order to lay the background for the quantization of this
    model, by introducing the form fields:
        \bea
            \tilde{A}&=&A_{i}J^{i}+C_{i}R^{i}  \nonumber \\
            \tilde{E} &=&E_{i}P^{i}+B_{i}Q^{i}  \label{bffields}
        \eea
    the BCEA model can be written as a BF theory associated with the Poincar\'{e}
    group:
        \be
            S[B,C,E,A]=\int_{M}tr(\tilde{E}\wedge F[\tilde{A}])
        \ee
    where $F[\tilde{A}]$ is the curvature of the Poincar\'{e} connection $\tilde{A}$,
    and has the expression:
        \be
            F[\tilde{A}]=d\tilde{A}+\frac{1}{2}[\tilde{A},\tilde{A}]=R_{i}[A]J^{i}+(DC_{i})R^{i}
        \ee
    with $R^{i}[A]$ and $DC^{i}$ given by (\ref{RA}) and (\ref{covder})
    respectively.
\bigskip

\section{The canonical structure of the BCEA theory}

\indent
    It was shown in the previous section that the BCEA model
    can be written as a Chern-Simons theory with the $I(ISO(2,1))$
    generalized connection ${\cal {A}} $, and therefore, its
    fundamental degrees of freedom are given by the
    holonomies of this connection along non-contractible loops in the
    spacetime $M$, modulo gauge transformations. The spacetime manifold
    under consideration has the topology $M=R\times \Sigma $, and we
    restrict ourselves to the case where the leaves $\Sigma $ of the
    foliation are 2-dimensional spacelike surfaces with
    the topology of a plane with one puncture. \\
\indent
    In this case, there is only one class of non-contractible
    loops. We can consider these to be loops at fixed time surrounding
    the puncture without any loss of generality. Consequently, the
    holonomy of the connection ${\cal {A}}$ along a loop $\gamma $
    representative of this class is given by the expression:
        \be
            W_{{\cal {A}}}(\gamma )={\cal {P}}\exp\:(\oint_{\gamma }{\cal {A}})\label{genhol}
        \ee
    where ${\cal {P}}$ in the r.h.s. of (\ref{genhol}) stands for the usual path
    ordering.\\
\indent
    Under these circumstances, it is straightforward to show
    that the explicit group structure of $I(ISO(2,1))$ is determined
    exclusively by the holonomy (\ref{genhol}). The generalized connection
    can be written as ${\cal A}=A+B+C+E$ with
        \be
            A=A_{i}J^{i},\;B=B_{i}Q^{i},\;C=C_{i}R^{i},\;E=E_{i}P^{i}.
        \ee
    and the holonomy of ${\cal A}$ can be computed by repeatedly using the
    identity
        \be
            W_{[{\cal {A}}_{1}+{\cal {A}}_{2}]}(t)=\;W_{\left[ W[{\cal {A}}_{1}]\,
                {\cal {A}}_{2}\,W^{-1}[{\cal {A}}_{1}]\right] }(t)
                W_{[{\cal {A}}_{1}]}(t)\label{wlsneq3}
        \ee
    where $W_{[A]}(t)\equiv {\cal P}\exp \{\int_{0}^{t}A(u)du\}$ is the unique
    solution of the differential equation:
        \be
            \frac{d}{dt}W_{[{\cal {A}}]}(t)=W_{[{\cal {A}}]}(t)A(t)\label{wlsnlneq}
        \ee
    with the initial condition $W_{[{\cal {A}}]}(0)=1$. A direct computation
    yields for the general form of the group element the expression
        \be
            W_{{\cal {A}}}(\gamma )\equiv G(g,\vec{a},\vec{b},\vec{c})=
            e^{a_{i}P^{i}+b_{i}Q^{i}+c_{i}R^{i}}g  \label{grpelem}
        \ee
    where the group parameters in (\ref{grpelem}) are given by
        \bea
            \vec{a} &=&\int_{0}^{1}\tilde{E_{\gamma }}(t)+\frac{1}{2}\int_{0}^{1}\int_{0}^{1}
            \epsilon(t-u)\tilde{B_{\gamma }}(u)\times \tilde{C_{\gamma }}(t)\,dudt \nonumber \\
            \vec{b} &=&\int_{0}^{1}\tilde{B_{\gamma }}(t)\\
            \vec{c} &=&\int_{0}^{1}\tilde{C_{\gamma }}(t) \nonumber\\
            g&=&{\cal {P}}exp\:(\oint_{\Gamma }{A})\nonumber\label{grparam}
        \eea
    and in (\ref{grparam}) we have used the notations
    $\tilde{E_{\gamma }}(t)\equiv W_{A_{\gamma }}(t)E_{\gamma }(t)W_{A_{\gamma }}^{-1}(t)$,
    $A_{\gamma }(t)\equiv A_{i}\frac{d\gamma^{i}}{dt}$, $\epsilon (t)$ for the
    sign function and $(\vec{a}\times \vec{b})^{i}=\epsilon^{ijk}\,a_{j}b_{k}$
    is the usual ``cross-product''.\\
\indent
    Once the explicit form of the group elements is known, one
    can immediately calculate the product of two group elements, the
    inverse of a group element, and the conjugate of a group element
    by another group element using the Baker-Campbell-Haussdorf formula.
    If $G_{1}$=$G(g_{1},\vec{a}_{1},\vec{b}_{1},\vec{c}_{1})$ and
    $G_{2}$=$G(g_{2},\vec{a}_{2},\vec{b}_{2},\vec{c}_{2})$ are two
    arbitrary group elements, we find that their product has the
    expression:
        \bea
            G_{1}G_{2}&=&G(g^{\prime },\vec{a^{\prime }},\vec{b^{\prime }},
                                    \vec{c^{\prime }})\nonumber\\
            g^{\prime }&=&g_{1}g_{2}  \nonumber \\
            \vec{a^{\prime }}&=&\vec{a}_{1}+(g_{1}\cdot \vec{a}_{2})+
                \frac{1}{2}\,\vec{b}_{1}\times (g_{1}\cdot \vec{c}_{2})+
                \frac{1}{2}\,\vec{c}_{1}\times(g_{1}\cdot \vec{b}_{2})\label{grpprod}\\
            \vec{b^{\prime }}&=&\vec{b}_{1}+(g_{1}\cdot \vec{b}_{2})\nonumber\\
            \vec{c^{\prime }}&=&\vec{c}_{1}+(g_{1}\cdot \vec{c}_{2})\nonumber
        \eea
    where $(g\cdot \vec{a})=g_{j}^{i}\,a^{j}$ with $g_{j}^{i}$ elements of the
    matrix of $g$ in the vector representation. By setting $G_{1}G_{2}$ equal to
    the identity element of the group in (\ref{grpprod}), it follows immediately
    that the inverse of an arbitrary group element $G(g,\vec{a},\vec{b},\vec{c}\,)$
    is given by the expression:
        \be
            G^{-1}(g,\vec{a},\vec{b},\vec{c}\,)=G(g^{-1},\;-g^{-1}\cdot\vec{a},
                    \;-g^{-1}\cdot \vec{b},\;-g^{-1}\cdot \vec{c}\,)\label{inv}
        \ee
\indent
    Finally, the conjugate of a group element $G(g,\vec{a},\vec{b},\vec{c}\,)$
    by another arbitrary group element $K(k,\vec{\alpha},\vec{\beta},\vec{\gamma}\,)$
    is given by:
        \bea
            KGK^{-1}&=&(kgk^{-1},\vec{A},\vec{B},\vec{C}\,)  \nonumber \\
            \vec{A}&=&(k\cdot \vec{a})+({\bf 1}-kgk^{-1})\cdot \vec{\alpha}+\frac{1}{2}
                \,\{[({\bf 1}+kgk^{-1})\cdot \vec{\beta}]\times (k\cdot \vec{c})+{}\nonumber \\
                &&{}+[({\bf 1}+kgk^{-1})\cdot \vec{\gamma}]\times (k\cdot \vec{b})
                +(kgk^{-1}\cdot \vec{\beta})\times \vec{\gamma}-\vec{\beta}\times
                (kgk^{-1}\cdot \vec{\gamma})\}  \nonumber \\
            \vec{B}&=&(k\cdot \vec{b})+({\bf 1}-kgk^{-1})\cdot \vec{\beta}\label{conj}\\
            \vec{C}&=&(k\cdot \vec{c})+({\bf 1}-kgk^{-1})\cdot \vec{\gamma}  \nonumber
        \eea
\indent
    Before proceeding with the explicit analysis of the    canonical structure of the BCEA theory, it is necessary to make a
    few remarks on the methods that will be used for this purpose. As
    mentioned earlier in Section 2, the fundamental degrees of freedom
    of the theory are given by the holonomies of the generalized
    connection ${\cal {A}}$ along non-contractible loops around the
    puncture. Since there is only one class of such loops, the
    physical configuration space will be labeled by the conjugacy
    classes of the $I(ISO(2,1))$ group which are now to be
    determined.\\
\indent
    Although it is possible in principle to analyze the
    conjugacy classes of a group directly on the group, for the
    general case the analysis is long and tedious. For Lie groups
    however, the analysis can be significantly simplified by
    restricting to the connected component of the group and reducing
    the determination of the conjugacy classes to the determination of
    the orbits of the action of the group in some vector
    representation (usually the adjoint or coadjoint representation)
    on its Lie algebra. If this latter orbit approach is used for the
    analysis of the canonical structure of the BCEA theory, the
    physical configuration space will be labeled correspondingly by
    the orbits of the group action on its algebra. In our particular
    case, as can be seen from (\ref{conj}), it is convenient to
    use a slightly different version of the orbit approach described
    above. Instead of working with a vector representation of the full
    group acting on its Lie algebra, we work on the group with the
    connected component of $SO(2,1)$, whose conjugacy classes are well
    known, and use only its vector representation to act on the remaining
    ideal of the algebra of $I(ISO(2,1))$.\\
\indent
    Having decided to use an orbit approach for labeling the
    physical configuration space raises a very important issue that
    must be considered, namely the classification of orbits. For any
    Lie group acting on some manifold, there is a well known result of
    the theory of invariants that states that there exists a certain
    number of algebraically independent invariant functions defined
    globally on the group that take constant values on the
    corresponding group orbits (and hence are independent of the gauge
    parameters). The number of such invariant functions depends on
    both the dimensions of the group and the manifold, and is finite
    if these dimensions are finite. Once again, although in principle
    it possible to determine such globally invariant functions on the
    group, in practice it is easier to work with the well-known
    Casimir invariants, which are related to the invariant functions
    on the group as follows. The Casimir invariants are elements of
    the enveloping algebra commuting with all the Lie algebra
    elements. Now, since any element of the Lie algebra can be viewed
    as a linear function on the Lie algebra if one uses an invariant
    scalar product, a Casimir invariant can be identified with a
    function on the Lie algebra that is invariant under
    the adjoint action, called a Casimir function. For instance if $%
    X=s^{i}J_{i}+a^{i}P_{i}+b^{i}Q_{i}+c^{i}R_{i}$ is a Lie algebra element, by
    using the invariant scalar product (\ref{scalprod}), the Lie algebra
    elements $P_{i}$ can be defined as the functions $P_{i}(X)=s_{i}$. The
    Casimir functions are therefore the germs of the invariant functions on the
    group, and hence, orbit invariants.\\
\indent
    Based on the above considerations, it is very natural to
    attempt to classify the orbits of the group action by means of the
    Casimir invariants or Casimir functions respectively. And in the
    case of compact Lie groups, such a classification is indeed
    possible since the Casimirs are in a one-to-one correspondence
    with the orbits of the action of the group on its Lie algebra. In
    the case of non-compact groups however, like $I(ISO(2,1))$, things
    become more complicated. In this case, the relation between
    Casimirs and orbits is not one-to one anymore, and while the
    Casimirs (if any) can still label some of the orbits, there are
    orbits, especially those of lower dimension, that have no
    corresponding Casimir (see for example \cite{Witten2}).
    Nevertheless, in particular cases, and $I(ISO(2,1))$ is one such
    case, the orbits of the group action can still be classified by
    means of invariants as follows. For orbits corresponding to
    Casimir invariants of the Lie algebra of the group, they are
    parametrized as usual by these invariants. For the orbits that do
    not correspond to Casimir invariants, one can still find either
    Casimir-like invariants for these orbits - invariants that do not
    commute with the Lie algebra in general case, but commute with the
    Lie algebra {\it on the orbit} - and/or scalar invariants, as the
    case may be (see below). In the following, we adopt this latter
    method and classify the orbits of the action of $I(ISO(2,1))$ on
    its Lie algebra using for this purpose both Casimir invariants
    and scalar invariants as necessity dictates.\\
\indent
    The independent Casimir invariants of the Lie algebra of
    $I(ISO(2,1))$ are relatively easy to determine. Following
    \cite{Abellanas}, it is straightforward to show that for the
    adjoint action of $I(ISO(2,1))$ on its Lie algebra there are only
    four algebraically independent Casimir invariants given by the
    expressions:
        \be
            C_{1}=J_{i}P^{i}+Q_{i}R^{i},\ C_{2}=P_{i}P^{i},\
            C_{3}=P_{i}Q^{i},\ C_{4}=P_{i}R^{i}\label{casimirs}
        \ee
    where $X=s^{i}J_{i}+a^{i}P_{i}+b^{i}Q_{i}+c^{i}R_{i}$ is a general
    Lie algebra element. As a direct consequence of the above number
    of invariants, the maximal dimension of the corresponding orbits
    is eight.\\
\indent
    We can now resume the analysis of the BCEA theory and
    proceed to the explicit determination of the adjoint orbits of the
    $I(ISO(2,1))$ group. By inspection of (\ref{conj}), the orbits can
    at once be separated into two major classes, depending on whether
    the group element $g$ is the identity element of $SO(2,1)$ or not.
    We investigate each of these cases separately.\\

\noindent
    \textbf{a. The case $g\neq{\bf 1}$}\\
    From the very beginning, this case can be separated into three
    distinct subcases, depending on whether $g$ is a rotation, a boost
    or a null transformation. The corresponding orbits are the most
    general orbits of the group action, having maximal dimensionality.
    It should be noted at this time that the sectors of the theory
    corresponding to these orbits are physically rather trivial in the
    sense that for such orbits one can always find a gauge in which the
    dynamics of the $B$, $C$ fields of the BCEA theory decouple from
    the dynamics of Einsteinian gravity.
\indent
    \begin{itemize}
        \item[\textbf{a1)}]
            If $g$ is a rotation, we can choose $k$ such that $%
            kgk^{-1}=exp\,(sJ_{0})$, and the vector gauge parameters $\vec{\alpha}$,
            $\vec{\beta}$, $\vec{\gamma}$ can be chosen such that the vectors
            $\vec{A}=ae_{0}$, $\vec{B}=be_{0}$, $\vec{C}=ce_{0}$ are all timelike
            and parallel to the axis of rotation of $kgk^{-1}$. The gauge orbit in
            this case is labeled by four real numbers $(s,a,b,c)$, corresponding to
            the four Casimir invariants:
                \be
                    P_{i}P^{i}=-s^{2},\ P_{i}J^{i}+Q_{i}R^{i}=-(sa+bc),\
                    P_{i}Q^{i}=-sc,\ P_{i}R^{i}=-sb \label{casinv}
                \ee
        \item[\textbf{a2)}]
            If $g$ is a boost (we only consider the connected
            component of  the boost subgroup), i.e. if we can choose $k$ such that
            $kgk^{-1}=exp\,(sJ_{1})$, the vector gauge parameters $\vec{\alpha}$,
            $\vec{\beta}$, $\vec{\gamma}$ can be chosen  such that the vectors
            $\vec{A}$, $\vec{B}$, $\vec{C}$ are pure  spacelike vectors. The
            gauge orbit in this case is labeled by  four real numbers $(s,a,b,c)$,
            corresponding to the four Casimir invariants:
                \be
                    P_{i}P^{i}= s^{2},\ P_{i}J^{i}+ Q_iR^i= as + bc,
                    \ P_{i}Q^{i}= sc,\ P_{i}R^{i}= sb
                \ee
        \item[\textbf{a3)}]
            If $g$ is a null transformation, (again, we only consider
            the connected component of the null subgroup) we can choose $k$ such that
            $kgk^{-1}=exp\{\,\frac{1}{2}\,(J_{0}+J_{1})\}$, and we can choose the
            vector gauge parameters $\vec{\alpha}$, $\vec{\beta}$, $\vec{\gamma}$ such
            that each of the vectors $\vec{A}$, $\vec{B}$, $\vec{C}$ is a null vector
            proportional to $(1,0,-1)$. The gauge orbits will labeled by three real
            numbers $(a,b,c)$, and the values of the independent Casimir invariants
            are given by all the possible combinations of the numbers:
                \be
                    P_{i}P^{i}= 0,\ J_{i}P^{i}+Q_iR^i= \pm a^{2},
                    \ Q_{i}P^{i}= \pm b^{2}\ R_{i}P^{i}= \pm c^{2}
                \ee
            In addition to the parameters $(a,b,c)$, the orbits will also be labeled by
            a set of discrete parameters describing the time
            orientation of the vectors $\vec{A}$, $\vec{B}$,
            $\vec{C}$, and consequently, these discrete parameters
            will introduce a degeneracy of the orbits relative to
            the values of the above Casimir invariants.
    \end{itemize}

\noindent
    \textbf{b. The case $g={\bf 1}$}\\
    In this case, the condition $g={\bf 1}$ implies that $P_{i}=0$,  and
    consequently the highest dimension of the corresponding orbits  is
    only six. The relevant equations in (\ref{conj}) reduce to:
        \bea
            \vec{A}&=&(k\cdot\vec{a})+\vec{\beta}\times(k\cdot\vec{c})+\vec{\gamma}
                        \times(k\cdot\vec{b})\nonumber\\
            \vec{B}&=&(k\cdot\vec{b})\label{conj1}\\
            \vec{C}&=&(k\cdot\vec{c})\nonumber
        \eea
    and from the particular form of the system (\ref{conj1}) it
    follows that there are two distinct cases.
\indent
    \begin{itemize}
        \item[\textbf{ b.1)}]
            If $\vec{B}\times \vec{C}\neq 0$, the vector gauge
            parameters $\vec{\beta}$, $\vec{\gamma}$ in (\ref{conj1}) can always be
            chosen such that $\vec{A}=0$. Using the remaining $SO(2,1)$ gauge freedom,
            one finds that the orbits can be parametrized by three real numbers $(a,b,c)$,
            corresponding to the scalar invariants:
                \be
                    b_{i}b^{i},\ b_{i}c^{i},\ c_{i}c^{i}\label{bcrossc1}
                \ee
            The value of these invariants depends upon whether the vectors $\vec{b}$,
            $\vec{c}$ are spacelike, timelike or null. Furthermore, the orbit
            also depends on discrete parameters specifying whether $\vec{b}$,
            $\vec{c}$ are future or past directed when they are timelike or
            null.\\
            The invariants (\ref{bcrossc1}) do not correspond to
            any of the Casimir invariants (\ref{casimirs}), but as mentioned
            earlier, for these orbits one can construct the Casimir-like
            quantities:
                \be
                    {Q}^{i}Q_{i},\,Q^{i}R_{i},\,R^{i}R_{i},
                \ee
            which commute with Lie algebra elements if the constraints $P_{i}=0$ are
            implemented.\\
            It is important to emphasize that this is the physically most interesting
            case, since for this configuration we cannot find a gauge where the $B,C$
            fields of the BCEA theory can be decoupled from gravity, i.e. a gauge where $%
            B\wedge C=0$.
        \item[\textbf{b.2)}]
            If $\vec{B}\times\vec{C}=0$, the vectors $\vec{B}$, $\vec{C}$
            are parallel, and in this case it is not  possible anymore to choose the
            vector gauge parameter $\vec{\beta}$, $\vec{\gamma}$ to cancel out the
            vector $\vec{A}$. However, they can be chosen such that they  cancel out the
            component of $\vec{A}$ that is  perpendicular to $\vec{B}$, $\vec{C}$, and
            consequently, by fixing $\vec{\beta}$, $\vec{\gamma}$,  we can choose
            without any loss of generality the  vector $\vec{A}$ to be parallel to the
            vectors $\vec{B}$, $\vec{C}$. Taking advantage once again of  the remaining $%
            SO(2,1)$ gauge freedom, the resulting  orbits are parametrized by three real
            numbers $(a,b,c)$, that correspond to the invariant scalars:
                \be
                    a_{i}b^{i},\ b_{i}b^{i},\ c_{i}c^{i}\label{bcrossc2}
                \ee
            Similar to the previous case, the expression of the  independent orbit
            invariants will depend upon whether the vectors $\vec{a}$, $\vec{b}$,
            $\vec{c}$ are timelike, spacelike or  null. Once again, in the timelike
            and null cases, the orbit will also  depend on discrete
            parameters labeling the time orientability of such vectors.
    \end{itemize}

\bigskip

\section{Examples of solutions}

\indent
    Once the distinct gauge orbits have been determined, the
    next logical step is to determine or identify solutions of the
    BCEA theory that will allow us to construct spacetime metrics
    corresponding to such orbits.\\

\subsection{\textbf{The point particle solution}}
\indent
    As it can be seen from the previous analysis, the most general
    gauge orbit is 8-dimensional and is characterized by four non-zero
    Casimir invariants. We restrict ourselves to the Casimir
    invariants (\ref{casinv}), i.e. to the case $({\bf a.1})$, and we
    proceed to construct a solution for the BCEA theory corresponding
    to this case.\\
\indent
    A set of $(B,C,E,A)$ fields compatible with the invariants (\ref{casinv})
    is given by:
        \bea
            E^{0}&=&ad\phi ;\;E^{1}=0;\;E^{2}=0  \nonumber \\
            A^{0}&=&sd\phi ;\;A^{1}=0;\;A^{2}=0  \nonumber \\
            B^{0}&=&bd\phi ;\;B^{1}=0;\;B^{2}=0  \label{flatpp} \\
            C^{0}&=&cd\phi ;\;C^{1}=0;\;C^{2}=0  \nonumber
        \eea
    and these fields are an obvious solution of the BCEA model since they
    identically satisfy the equations of motion (\ref{eqnmotn}). In this form
    however, the matrix of the triad form fields is singular, and therefore one
    cannot directly construct a spacetime metric using the co-triads in (\ref%
    {flatpp}).\\
\indent
    In order to overcome this difficulty, we will use a slight
    variation of the method described in \cite{sousagerb}. By
    setting $s=M$, $a=J$ in (\ref{flatpp}), the resulting triad
    and connection form fields:
        \bea
            E^{0}&=&Jd\phi ;\;E^{1}=0;\;E^{2}=0\nonumber\\
            A^{0}&=&Md\phi ;\;A^{1}=0;\;A^{2}=0\label{flatpp1}
        \eea
    are identical to the triad and connection form fields $(3.14)$
    in \cite{sousagerb} for a point-particle of mass $M$ and spin $J$
    (which, by an abuse of language, will subsequently be called a flat
    point-particle) in pure Einsteinian gravity \cite{sousagerb},
    \cite{dgh}. Consequently, the BCEA model admits a flat point-particle
    solution given by (\ref{flatpp}) with (\ref{flatpp1}), and this solution
    corresponds to the most general gauge orbit under consideration.\\
\indent
    Since invertibility of the triad form fields is a gauge
    dependent property, constructing a solution with invertible
    triad form fields from (\ref{flatpp}),(\ref{flatpp1}) is a
    matter of straightforward calculation. As mentioned earlier,
    we will use for this purpose an approach similar to that
    used in \cite{sousagerb}, the only difference being that
    instead of using a representation of the $I(ISO(2,1))$ Lie
    algebra generators, we use the gauge transformations
    (\ref{eqngauge}).\\
\indent
    A simple gauge transformation that yields invertible triad
    form fields is given by:
        \be
            \beta^{0}=t;\;\beta^{1}=\frac{r\cos\phi}{1-M};
            \;\beta^{2}=\frac{r\sin\phi}{1-M}
        \ee
    with all remaining gauge parameters zero. The resulting gauge
    transformed triad fields have the expression:
        \bea
            E^{0}&=&dt+Jd\phi\nonumber\\
            E^{1}&=&\frac{1}{1-M}\,\cos \phi dr-r\sin \phi d\phi\\
            E^{2}&=&\frac{1}{1-M}\,\sin \phi dr+r\cos \phi d\phi \nonumber\label{flatpp2}
        \eea
    while the rest of the form fields in (\ref{flatpp}), (\ref{flatpp1}) remain
    unaffected by this gauge transformation. It is straightforward to show that
    these gauge transformed fields are a solution of the BCEA theory. Finally,
    the triad form fields (\ref{flatpp2}) yield the familiar flat point-particle
    metric \cite{sousagerb}:
        \be
            ds^{2}=-(dt+Jd\phi )^{2}+\frac{dr^{2}}{(1-M)^{2}}+r^{2}d\phi ^{2}
        \ee
    The fact that we can recover a flat space solution should not come as a
    surprise since our analysis shows that in the case where $A$ is not trivial,
    i.e. for $g\neq 1$, we can always chose a gauge where $\vec{B}\times \vec{C}=0$.
    In this gauge the dynamics of the $B,C$ fields decouples from the
    dynamics of 2+1 geometry as it can be seen from (\ref{eqnmotn}).

\subsection{\textbf{The BTZ black-hole}}
\indent
    The geometry which realizes the orbit $({\bf b.1})$ is the
    BTZ black-hole \cite{btz1}, \cite{btz2}. The BTZ solution for the
    BCEA theory with $A^{0}=A^{1}=A^{2}=0$ is given by the fields
    \cite{bceabtz,cangemi}:
        \bea
            E^{0}&=&2\sqrt{\nu^{2}(r)-1}(\frac{r_{+}}{l}dt-r_{-}d\phi)\nonumber\\
            E^{1}&=&\frac{2l}{\nu(r)}d[\sqrt{\nu^{2}(r)-1}]\label{btztriad}\\
            E^{2}&=&2\nu(r)(-\frac{r_{-}}{l}dt+r_{+}d\phi)\nonumber\\
            B^{0}&=&\frac{r_{-}}{l}dt-r_{+}d\phi\nonumber\\
            B^{1}&=& -ld[\nu(r)+\sqrt{\nu^{2}(r)-1}]\nonumber\\
            B^{2}&=& \frac{r_{+}}{l}dt-r_{-}d\phi\nonumber\\
            C^{0}&=& -\frac{r_{-}}{l^{2}}dt+\frac{r_{+}}{l}d\phi\label{bcflds}\\
            C^{1}&=& d[\nu(r)-\sqrt{\nu^{2}(r)-1}]\nonumber\\
            C^{2}&=& \frac{r_{+}}{l^{2}}dt-\frac{r_{-}}{l}d\phi\nonumber
        \eea
    where
        \bea
            r^{2}_{+}&=&\frac{Ml^2}{2}\,\{1+\sqrt{1-(J/Ml)^{2}} \}\nonumber\\
            r^2_-&=&\frac{Ml^2}{2}\,\{1-\sqrt{1-(J/Ml)^{2}}\}\label{horradii}
        \eea
    are the outer and respectively inner horizon radii, satisfying $r_+r_-= Jl/2$
    and the function $\nu(r)$ is given by the expression:
        \be
            \nu^{2}(r)=\frac{r^{2}-r^{2}_{-}}{r^{2}_{+}-r^{2}_{-}}
        \ee
    The parameters $M$ and $J$ in (\ref{horradii}) are the  quasilocal mass and
    angular momentum of the black-hole, $l$ is related  to the cosmological
    constant through the relation:
        \be
            \Lambda=-\frac{1}{l^{2}}<0
        \ee
\indent
    In order to calculate the holonomy of the generalized connection $%
    {\cal{A}}$, we choose as a loop a circle of radius $r_{+}$ at constant
    time. In this case, the calculation of the  holonomy for the BTZ fields
    (\ref{btztriad}) and (\ref{bcflds}) is straightforward, and we obtain
    \footnote[4]{In the  expression of the holonomy we have dropped, for simplicity
    reasons, an overall factor  of $2\pi$ arising from the integral over $\phi$.}:
        \be
            W_{[\cal{A}]}^{BTZ}(\gamma)= exp\:[2r_{+}P^{2}+(r_{+}Q^{0}-r_{-}Q^{2})-
                            \frac{1}{l}\;(r_{+}R^{0}+r_{-}R^{2})]\label{bhhol}
        \ee
\indent
    The vectors $\vec{a}$, $\vec{b}$ ,$\vec{c}$ corresponding
    to the holonomy (\ref{bhhol}) have the components:
        \bea
            \vec{a}&=&(0,0,2r_{+})\nonumber\\
            \vec{b}&=&(r_{+},0,-r_{-})\label{holvec}\\
            \vec{c}&=&-\frac{1}{l}\;(r_{+},0,r_{-})\nonumber
        \eea
    and we can clearly see that this corresponds to the case ($\textbf{b.1}$)
    where $g=1$ and $\vec{b}$, $\vec{c}$ are not parallel. In this case,
    the orbit invariants are $b_{i}b^{i}$, $b_{i}c^{i}$, $c_{i}c^{i}$ with
    $\vec{b}$, $\vec{c}$ being future and past timelike vectors.
    Using (\ref{horradii}), the explicit forms of the invariants in
    terms of the BTZ black-hole parameters $M$, $J$, $l$ are given by:
        \bea
            a\equiv b_{i}b^{i}&=&-[M^{2}-(J/l)^{2}]^{1/2}\nonumber\\
            b\equiv b_{i}c^{i}&=&lM \label{btzcas}\\
            c\equiv c_{i}c^{i}&=&-l^{2}[M^{2}-(J/l)^{2}]^{1/2}\nonumber
        \eea
    and consequently, the gauge orbit corresponding to the BTZ  black-hole
    solution of the BCEA theory is parametrized by  three real parameters $(a,b,c)$.
    Note that the curvature $l$ of the spacetime is a constant of integration
    and so appears naturally as a dynamical parameter, implying that the
    parameters $(a,b,c)$ are all independent. This is in contrast to
    the pure Einstein case, where the  curvature of the spacetime occurs as a
    fixed parameter of the  action.

\bigskip

\section{Discussion}

\indent\
    In this paper we have studied the phase space structure of
    2+1 gravity coupled to a pair of topological matter fields
    $B,C$. Using the formulation of this theory as an $I(ISO(2,1))$
    Chern-Simons model, we have identified the different sectors of
    this theory and have constructed or determined corresponding
    geometries of physical interest. Among the different sectors, two
    different types of solution emerge as relevant. In the first type,
    the dynamics $B,C$ fields can be decoupled from the dynamics of
    the geometry, and the model is equivalent to (2+1)-dimensional
    gravity in flat space on which form fields are superimposed. In
    the second type of solution the $B,C$ fields cannot be decoupled
    from gravity anymore. The dynamics of the fields and of the
    geometry are strongly interrelated, as one would expect from a
    theory where gravity is coupled to matter. Illustrative of this
    case is the BTZ black-hole solution, with the surprising result
    that the dynamical parameters of the solution include besides the
    mass and angular momentum of the black-hole - the parameters of
    the traditional (2+1)-dimensional theory of gravity with
    cosmological constant - the cosmological constant itself.\\
\indent
    Now that we have a clear picture of the classical dynamics of the BCEA model,
    we will need to address its quantization. Since the BTZ black hole is
    a solution of this model, understanding its quantization should allow us to
    give a full description of a quantum black hole in the presence of matter
    fields, a question that so far has never been addressed. Several strategies
    can be deployed in this direction. Since this theory can be formulated as a
    $I(ISO(2,1))$ Chern-Simons model we can first perform a Chern-Simons
    quantization of the model using the description of the classical phase space
    given here. As we have seen, the theory can also be formulated as a Poincar%
    \'{e} $BF$ theory which opens the way to a spin foam quantization of the
    model \cite{LK}. The most challenging issue in this latter approach would be
    to identify at the quantum level the different sectors of the classical
    theory. We hope to return to these issues in the near future.\\

\bigskip

\noindent
    \textbf{Acknowledgments}\\
\indent
    This work was supported in part by the Natural Sciences
    \& Engineering Research Council of Canada. Two of the
    authors, R.B.M and E.M.P would also like to thank M. Reisenberger
    for many useful discussions during his visits at the Perimeter
    Institute for Theoretical Physics.

\end{document}